   \documentclass[12pt,preprint]{aastex}

\begin{document}

\title{Spectroscopic Analysis of an EIT Wave/Dimming Observed by Hinode/EIS}

\author{F. Chen, M. D. Ding, and P. F. Chen}
\affil{Department of Astronomy, Nanjing University, Nanjing 210093, China}
\affil{Key Laboratory for Modern Astronomy and Astrophysics (Nanjing
University), Ministry of Education, Nanjing 210093, China}
\email{dmd@nju.edu.cn}


\begin{abstract}
EIT waves are a wavelike phenomenon propagating outward from the
coronal mass ejection (CME) source region, with expanding dimmings
following behind. We present a spectroscopic study of an
EIT wave/dimming event observed by Hinode/EIS. Although the identification of
the wave front is somewhat affected by the pre-existing loop structures, the expanding
dimming is well defined. We investigate the line
intensity, width, and Doppler velocity for 4 EUV lines. In addition to
the significant blue shift implying plasma outflows in the dimming
region as revealed in previous studies, we find that the widths
of all the 4 spectral lines increase at the outer edge of the dimmings.
We illustrate that this feature
can be well explained by the field line stretching model, which claims
that EIT waves are apparently moving brightenings that are generated by
the successive stretching of the closed field lines.
\end{abstract}


\section{Introduction}
EIT waves were first observed by the EUV Imaging Telescope (EIT)
aboard the {\it Solar and Heliospheric Observatory} ({\it SOHO}) \citep{mos97,
tho98}. They are best seen in the running difference images as bright
fronts with a propagation speed of $<500$ hundred km s$^{-1}$, followed
by an expanding dimming region \citep{tho98}. More
properties of EIT waves were presented by \citet{del99}, \citet{kla00},
and \citet{tho09}. EIT waves can be observed at several wavelengths,
such as 175 \AA, 195 \AA, 284 \AA, and 304 \AA~\citep{wil99,zhu04,lon08}.
Regarding the relationship between EIT waves and coronal mass ejections
(CMEs), \citet{pat09} claim that the EIT wave front is outside the CME
frontal loop, whereas \citet{chen09} and \citet{dai10} claim that the
EIT wave front is cospatial with the white-light frontal loop of CMEs.

Moreton waves are another kind of wavelike phenomenon seen in the
chromosphere. They are seen as arclike H$\alpha$ disturbances
propagating to a large distance away from the eruption site, with
velocities ranging from 500 to 2000 km s$^{-1}$ \citep{mor60}. It was
suggested that the footprints of a coronal fast-mode wave or shock wave
sweeping the chromosphere, as the wave moves through the tenuous corona,
would produce the apparent propagation of H$\alpha$ disturbances
\citep{uchi68}. This implies the existence of a fast-mode shock wave in
the corona, although it was not observed directly at that time.

After EIT waves were discovered, they were first suggested to be the
coronal counter-parts of Moreton waves. Therefore, according to
Uchida's model, EIT waves were ever considered to be a fast-mode MHD
wave. For example, \citet{wang00} found that the ray path of fast-mode
waves matches the propagation of the EIT wave fronts, as well as
reproducing their tendency of avoiding active regions and coronal holes.
\citet{wu01} performed a 3-D numerical simulation of the perturbation
resulting from a pressure pulse. Their result showed that the fast-mode
wave front could reproduce many properties of the observed EIT waves.
More events were analyzed by \citet{war01,war04}, who found that similar
propagating features appear in the coronal and chromospheric spectral
lines and suggested that they are signatures of the same physical
disturbance, i.e., a freely propagating fast-mode MHD shock. However, an
obvious problem of the fast-mode wave explanation is that the speeds of
EIT waves are statistically $>3$ times smaller than those of Moreton
waves. \citet{wu01} and \citet{war04} noticed that Moreton waves are
visible only near the flare sites, while
EIT waves are mostly observed at larger distances. Therefore, they
suggested that the speed difference could be a result of deceleration
during the wave propagation, which, however, was not supported by
\citet{eto02}. One the other hand, considering the behaviors of EIT
waves when they encounter magnetic separatrices, \citet{del00} suggested
that the bright front may be a result of compression of coronal plasma,
caused by the interaction between CME-induced expansion of magnetic
field lines and surrounding field lines.
\citet{chen02,chen05} performed numerical simulations of the MHD
process of the CME-induced perturbation and found a piston-driven shock
running ahead and a slower moving wavelike structure following
behind. They proposed that the former corresponds to the coronal
Moreton waves while the latter to the EIT waves, respectively. This
model was supported by \citet{har03}, who presented a TRACE
observation of a fast moving ``weak wave" with almost no
line-of-sight (LOS) mass motion and a slower moving ``bright wave"
followed by prominent dimmings.

Spectroscopic observations can provide us more insight into the physical
nature of EIT waves. \citet{asa08} presented a spectroscopic
observation of an MHD fast mode shock wave visible in soft X-rays.
Unfortunately, some of the EIT images suffered from scattered light in
the telescope; therefore, the EIT wave front was unclear. In this paper,
we present a Hinode/EIS observation of an EIT wave event. We obtain not
only the high-resolution Doppler velocity in the region that the wave
passes through, but also the line width distribution for the first time.
We describe the observation and data analysis in \S2; our results are
shown in \S3, followed by discussions about the mechanism of line
broadening and its implication to the model of EIT waves in \S4.

\section{Observations and Data Analysis}
The EIT wave event that we study here occurred in the active region NOAA
10956 on 2007 May 19.  It was associated with a B9 flare and a CME. The
SOHO/EIT was in CCD bakeout during the event, while the Extreme
UltraViolet Imager (EUVI) on board the Solar TErrestrial RElations
Observatory (STEREO; Howard et al. \citeyear{how08}) observed the event
in four wavebands. This event, in particular the propagation speed, has
been investigated by \citet{lon08} and \citet{ver08}.

The event was also observed by the Extreme-ultraviolet Imaging
Spectrometer (EIS). The EIS observation of NOAA 10956 started at
11:41:23 UT and ended at 16:35:03 UT using the 1\arcsec~slit that
rastered across the active region with a step of 1\arcsec~and an
exposure time of 40 s. The time gap is around 13.5 s between successive
exposures. The field-of-view (FOV) is 330\arcsec~in the raster
direction and 304\arcsec~in the slit direction. The EIS observation
covered the EIT wave event. It thus provides us a good opportunity to
make a spectroscopic study of the EIT wave and the ensuing dimmings.

The details of the EIS instrument are described in \citet{cul07}.
The two EIS detectors cover the wavelength ranges 170--210 \AA~and
250--290 \AA, therefore providing us spectral lines in a wide range
of emission temperatures. The EIS has 1\arcsec, 2\arcsec~slits, and
40\arcsec, 260\arcsec~slots available. The slit raster obtains high
spectral resolution data, while the slot observation provides
transition region and coronal monochromatic images.

We process the raw data using eis$\_$prep.pro in SSW. This process
flags bad data points and converts DNs to physical units (ergs
cm$^{-2}$ s$^{-1}$ sr$^{-1}$ \AA$^{-1}$ ). For this particular
study, we select several coronal lines with a broad range of
temperatures, as listed in Table 1. The \ion{Fe}{14} $\lambda$274.20 line is
blended with the \ion{Si}{7} $\lambda$274.18 line, while this
contamination can be estimated by the ratio of the \ion{Si}{7}
$\lambda$274.18 and $\lambda$275.35 lines. In this observation, we find
that the intensity of the \ion{Si}{7} $\lambda$275.35 line is rather weak,
thus the \ion{Si}{7} $\lambda$274.18 line is almost mixed with the
noise of the background. Therefore, we ignore the blending as
\citet{you07} suggested. Finally, we make Gaussian fittings to all the
lines using mpfitexpr.pro in SSW, with a fitting range of $\pm$0.2
\AA~from the default line center.

\begin{figure}
\epsscale{1}
\plotone{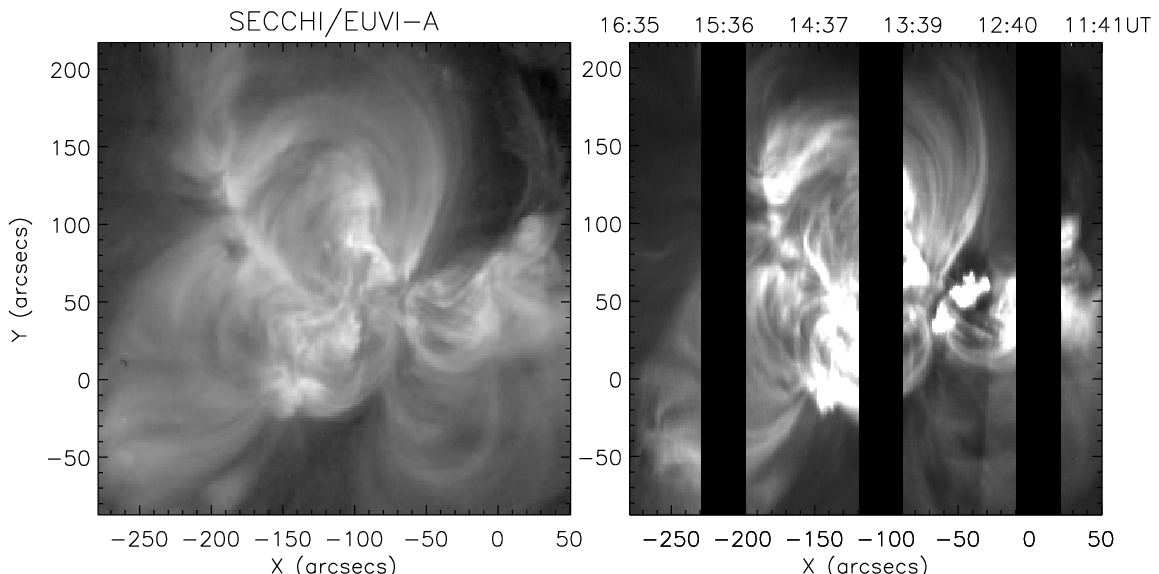}
\caption{SECCHI/EVUI-A 195 \AA~image at 12:32:00
UT and EIS \ion{Fe}{12} $\lambda$195.12 image
from 11:41:23 UT to 16:35:03 UT.}\label{fig1}
\end{figure}

To study the wave phenomena, we should coalign the EIS observation with
the STEREO-A/EUVI observation. We calculate the correlation
between the \ion{Fe}{12} 195 \AA~image from EUVI-A at 12:32 UT and the
intensity map obtained from EIS for the same emission line, using a
50\arcsec$\times$50\arcsec~box in order to get the offsets. Figure
\ref{fig1} shows the EUVI and EIS data after coalignment, where their
correlation coefficient is as high as $\sim 95\%$. Since STEREO
A and Hinode/EIS view the Sun from different angles, the coalignment of
optical-thin EUV images should take the stratified structure into
account, which involves the tomography technique using both STEREO A and B
images. Fortunately, STEREO A and Hinode were separated by only $\sim
4^\circ$ on 2007 May 19, which makes the direct translation being
sufficient as indicated by the high correlation coefficient.

\begin{figure}
\epsscale{1}
\plotone{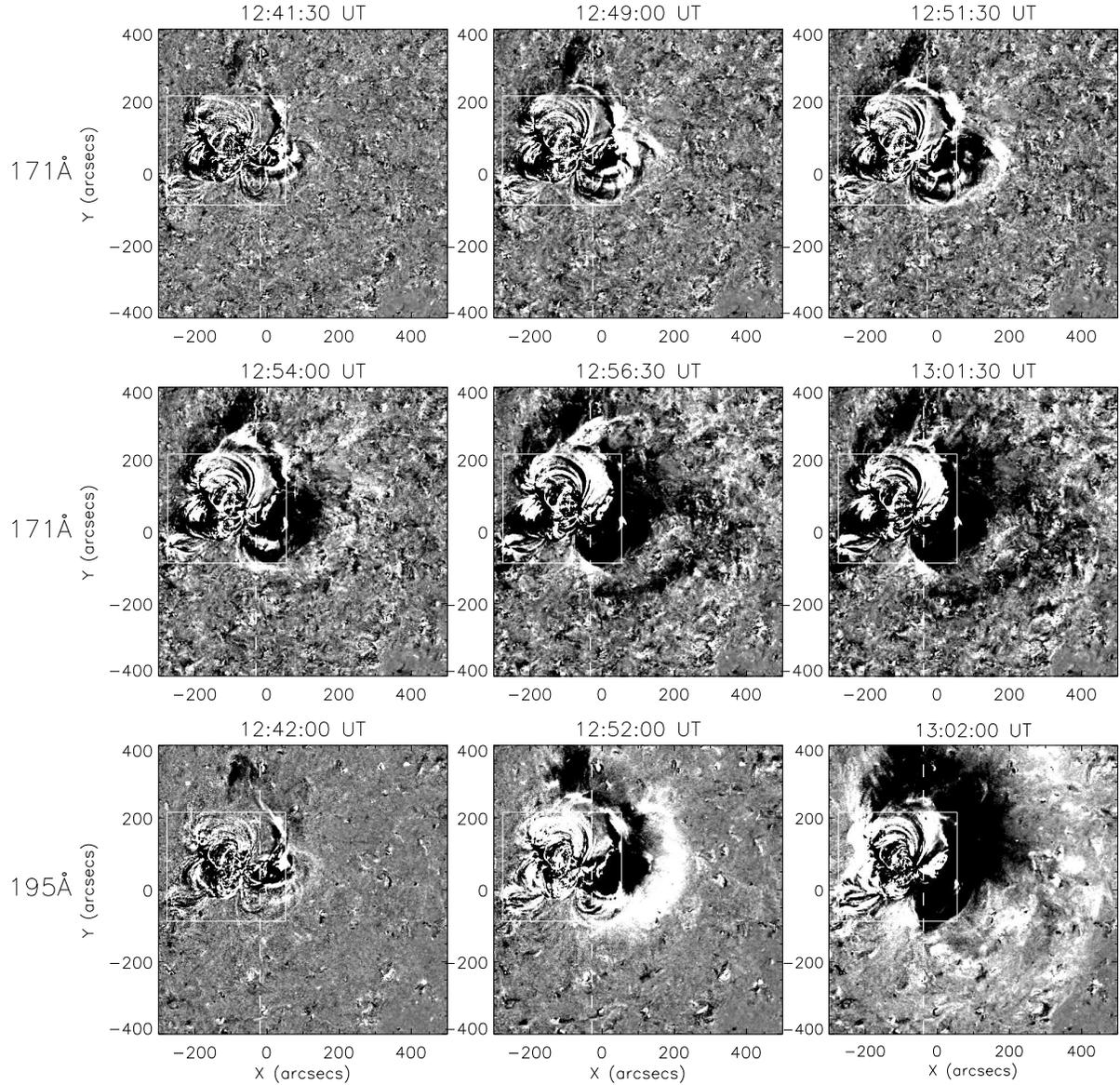}
\caption{Base difference images of SECCHI/EUVI-A
171 \AA~(top and middle rows) and 195 \AA~(bottom row).
The images at 171 \AA~are substracted by the image at
12:31 UT and the images at 195 \AA~are subtracted by
the image at 12:32 UT, with the solar rotation corrected.
The white boxes indicate the FOV
of EIS. The dashed vertical lines represent
the position of the silt at the time of each panel.}\label{fig2}
\end{figure}

For this event, \citet{lon08} found that the EIT wave front propagated
at a low velocity initially and significantly accelerated at around
12:50 UT in high-cadence SECCHI/EUVI-A 171 \AA~data (see the left column
of Fig. 3 in their paper). Figure \ref{fig2} shows the relation between
the EIS scan and the EIT wave seen from the base difference images of
SECCHI-A 171 \AA~and 195 \AA. The arc-like brightening was first
seen at 12:41 UT in the 171 \AA~image and at 12:42 UT in the 195 \AA~
image, while it was most clearly identified after 12:51 UT.
Unfortunately, the leading edge of the wave front was out of the EIS FOV
at that time, with only part of the brightening still lying in the EIS
FOV. The whole wave front propagated out of the EIS FOV after 12:54 UT.
The EIS raw data recorded the observation time of all rasters;
therefore, we can identify the time and position of the rasters when
they encountered the EIT wave. We find that the slit positions from
$x \sim -21$\arcsec~to $x \sim -27$\arcsec~and those from $x \sim
-28$\arcsec~to $x \sim -34$\arcsec~correspond to the initial and
acceleration stages of the EIT wave, respectively. It is
possible that some of the brightening seen near the active region during the
early time may be due to the displacement of pre-existing loop
structures. In fact, at least to some extent, the coronal loops can
affect the identification of the wave front and the measurement of
its width.

\section{Results}

\begin{figure}
\epsscale{1}
\plotone{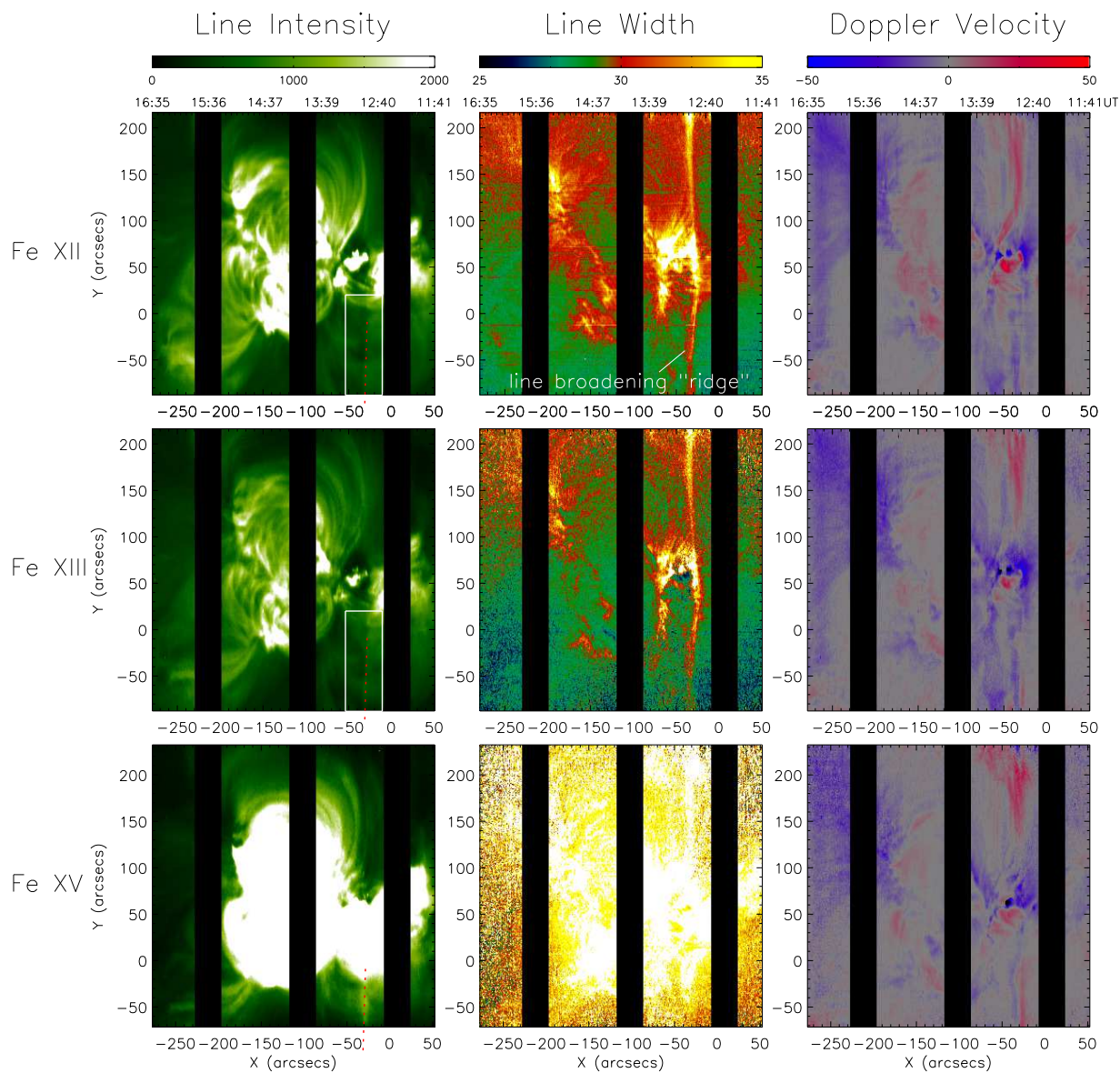}
\caption{Line intensity (left), width (middle), and
Doppler velocity (right) from 11:41:23 UT to
16:35:03 UT for 3 coronal lines. From top to bottom,
the three rows are for the \ion{Fe}{12} $\lambda$195.12,
\ion{Fe}{13} $\lambda$202.04, and \ion{Fe}{15} $\lambda$284.16
lines, respectively. The red dotted line is the same as that in Figure \ref{fig4}.
The white boxes indicate the region which is studied in detail below. Note that, the FOV of the \ion{Fe}{15} image
is actually different from that of the \ion{Fe}{12} and \ion{Fe}{13} images, because
of the offset between the two CCDs of EIS.}\label{fig3}
\end{figure}

After processing the coaligned data sets, we analyze in detail the EIT
wave on 2007 May 19 using the EIS observations. We obtain the line
intensity, line width, and Doppler velocity for several lines with
different formation temperatures. Figure \ref{fig3} shows the line
intensity, line width, and Doppler velocity for the \ion{Fe}{12}
$\lambda$195.12, \ion{Fe}{13} $\lambda$202.04, and \ion{Fe}{15}
$\lambda$284.16 lines. Figure \ref{fig4} presents the results of the
region of interest (a zoomed view of the white box in Figure
\ref{fig3}). We also analyze the EIS observation of the same active
region at about 3 hours before the eruption, with the results shown in
Figure \ref{fig5}.

We measure the positions of the wave front from 12:42 to 12:52
UT using the 195 \AA~data and those from 12:46 to 12:54 UT using the 171
\AA~data. The vertical bars in Figure \ref{fig4} indicate the width of
the bright structure and its propagation during a single EIS exposure;
the horizontal ones indicate the accuracy of EIS--EUVI coalignment,
i.e., $\sim 2\arcsec$. Note that the red bar at 12:54 UT is completely
out of the EIS FOV.

\subsection{The Intensity Decrease}
\begin{figure}
\epsscale{1}
\plotone{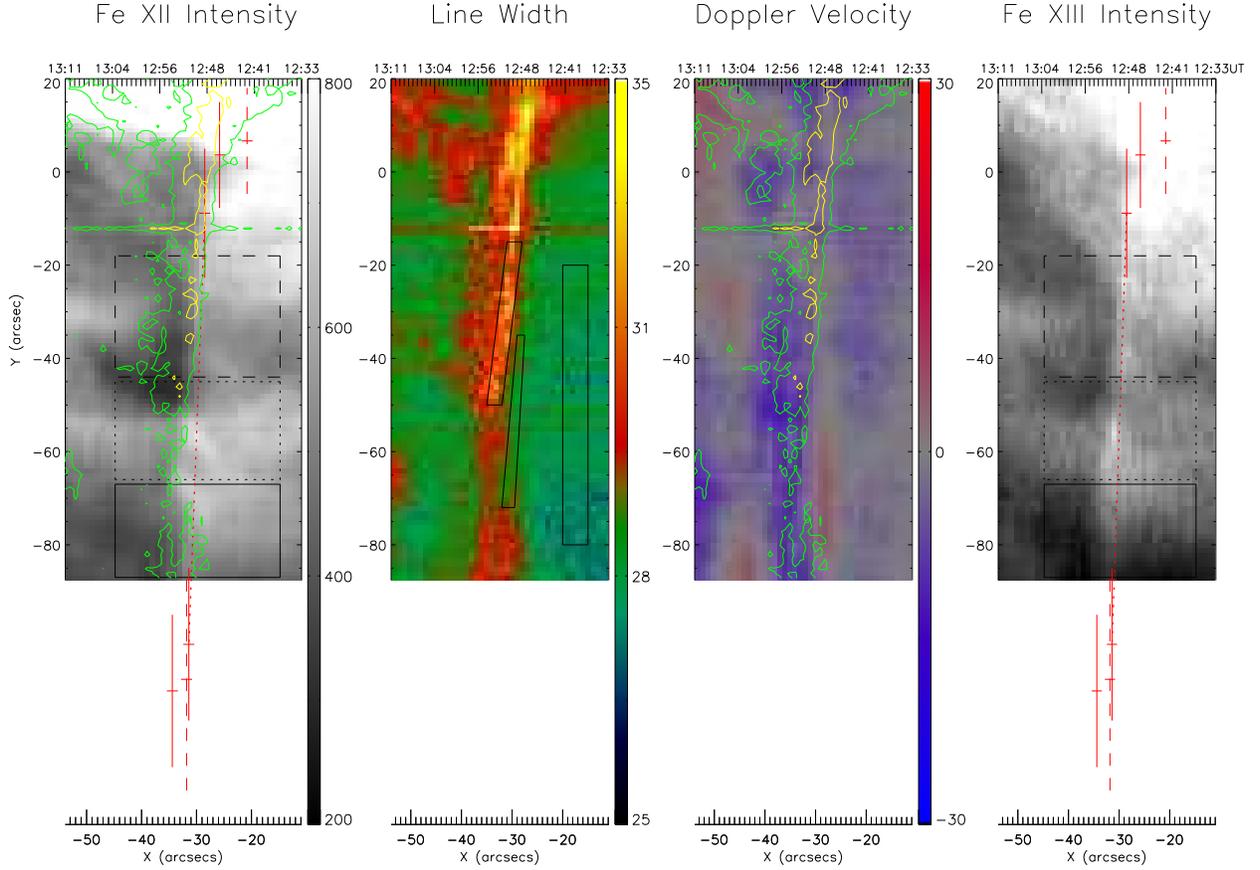}
\caption{Zoomed view of the line broadening ``ridge",
an enlargement of the white box shown in Figure \ref{fig3}. The left 3 panels show the line intensity,
width, and Doppler velocity for the \ion{Fe}{12} $\lambda$195 line
and the right panel shows the
intensity for the \ion{Fe}{13} $\lambda$202.04 line. The red solid and
dashed vertical bars in the first and forth panels indicate the wave front measured
from the EUVI-A 171 and 195 \AA~images, respectively. The middle points of the bars at
12:49 and 12:51 UT measured from the 171 \AA~image are connected by the
red dotted line that corresponds to the wave front propagation during the observation gap.  The intensity
and LOS velocity for the \ion{Fe}{12} line are overlaid by the
contours of the line width, with contour levels of 30 m\AA~(green)
and 32 m\AA~(yellow). The rectangles in the first and fourth panels indicate
the area selected for calculation of the intensity decrease.
The dashed, dotted, and solid lines
are for Y positions of $-18$\arcsec~to $-44$\arcsec,
$-45$\arcsec~to $-66$\arcsec~and $-67$\arcsec~to $-87$\arcsec~,
respectively. The quadrilaterals
in the second panel indicate the areas selected for
calculation of the line width.}\label{fig4}
\end{figure}

\begin{figure}
\epsscale{1}
\plotone{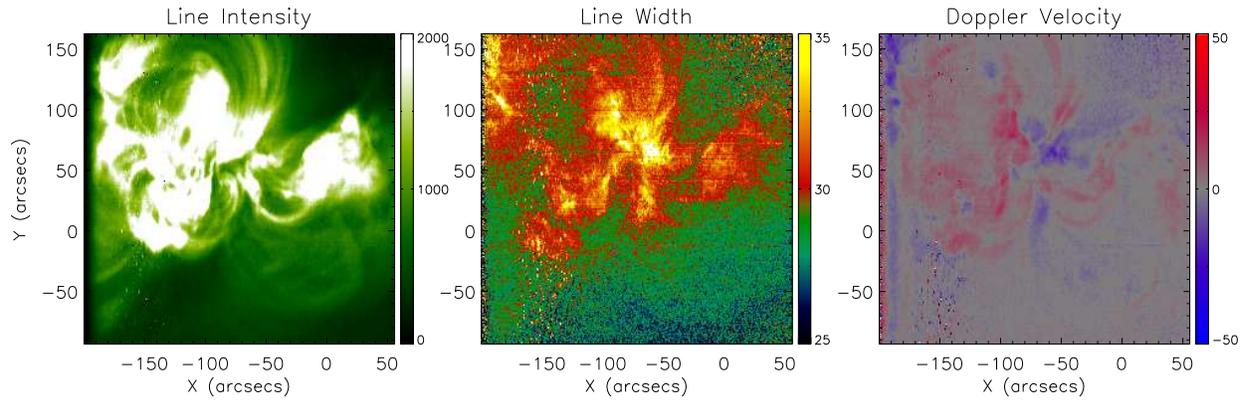}
\caption{Line intensity, width, and Doppler velocity maps
for the \ion{Fe}{12} $\lambda$195.12 line for the
active region before the eruption. The scanning time is from
09:42:12 UT to 10:31:24 UT. There is no significant
structure in the southern part of the active region,
in particular in the width and velocity maps.}\label{fig5}
\end{figure}

We compose the 2-dimensional images for the 3 EUV lines from each scan
in the left column of Figure \ref{fig3}, although the emissions are not
simultaneous in the E--W direction. Note that we have adjusted the color
scale to illustrate more clearly the intensity variation so that the
core of the active region is saturated. In contrast to the STEREO/EUVI-A
image (shown in Figure \ref{fig1}, left panel), the intensity in the
region to the east of $x \sim -30$\arcsec~is significantly lower than
that in the region to the west. The boundary of intensity decrease is
inclined to the slit direction by a small angle, because the slit was
moving along the W--E direction while the dimming region
expanded southward. Furthermore, it is just to the left of (i.e.,
behind) the red dotted line, an indicator of the wave front propagation,
as shown in Figure \ref{fig4}. Therefore, this region of intensity decrease
in the EIS map is considered to be the dimming region associated with
the EIT wave. The left panel of Figure \ref{fig5} shows that there is no
intensity decrease near $x \sim -30$\arcsec; therefore, the intensity
decrease during the eruption should not be a feature of the active
region itself, such as a coronal hole.

By investigating the intensity maps of different lines,
we find that the intensity decrease is prominent in the
\ion{Fe}{12} and \ion{Fe}{13} lines, while it is not evident
in the \ion{Fe}{14} and \ion{Fe}{15}
lines as well as the transition region lines like
\ion{He}{2} $\lambda$256.32. From Figure \ref{fig4}, one can
find a brightening feature just along the red dashed line in \ion{Fe}{13}.
It is supposed to be the bright wave front. The low contrast of the brightening
feature reflects a systematic drawback of observing fast propagating structures
using a long exposure time and a low raster cadence of the EIS. The propagation
distance of the wave front during
a single EIS exposure, which is at least 10\arcsec~(depending on the
wave propagation speed), is roughly comparable with the width
of the wave front seen in the 171 \AA~images. This makes the wave front look
very fuzzy in the figure reconstructed from the scanning observation. For this event,
the wave front is also seen in the 284 \AA~image of EUVI,
which is wider than that seen in the 171 and 195 \AA~images, especially
in the southern part (Fig. 1 in \citealt{lon08}). By comparison, the wave front seen in the image of
the \ion{Fe}{15} line of EIS is the clearest.  Furthermore, we find that the image contrast of EIS,
defined as the ratio of the maximum value of the active region to the average value over a
quiescent region, is lower than that of EUVI, since the quiescent value of the EIS
image is much higher than that of the EUVI, while the maximum values are not significantly
different. This can explain the lower contrast of the wave front seen in the EIS images
than that in the EUVI images.

\begin{figure}
\epsscale{1}
\plotone{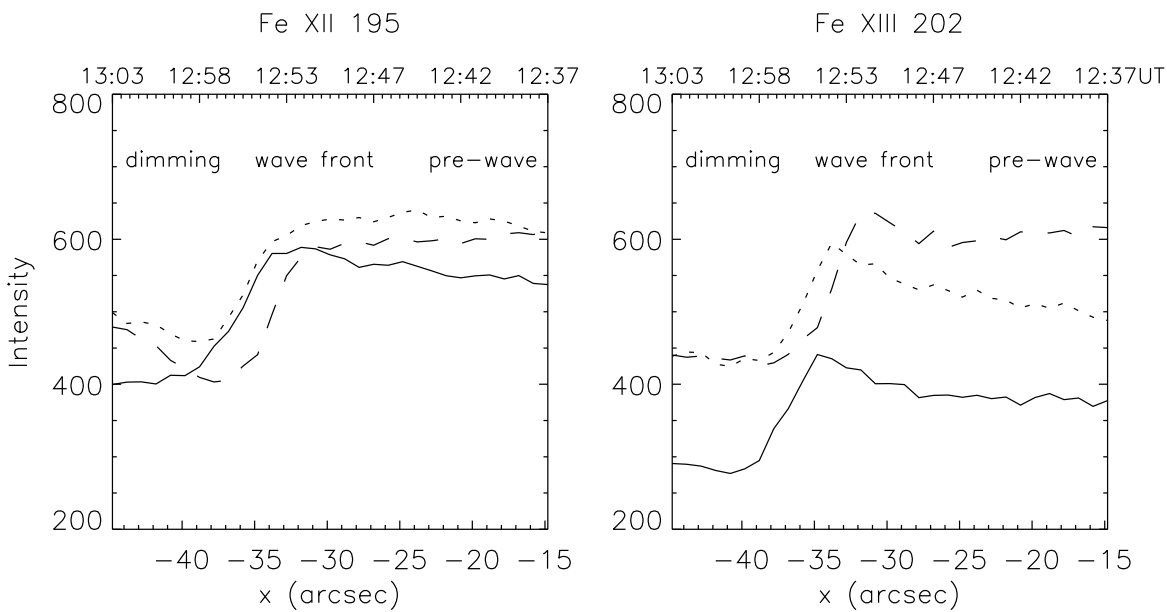}
\caption{Average intensity over the Y-direction for the three
areas indicated in Figure \ref{fig4} (first and fourth panels),
showing the intensity variation across the boarder of the dimming and
the wave front. The line styles are as the same as those in Figure
\ref{fig4}. The dash line corresponds to the area close to the source
region of the eruption, the dotted line to an area in the middle, and
the solid line to an area further away. The local intensity
minimum for the three lines moves to the left sequentially, implying
the propagation of the intersection between the expanding dimming and
the slit. In comparison, the wave front for the \ion{Fe}{12} line is
not as evident as that for the \ion{Fe}{13} line.}\label{fig6}
\end{figure}

We calculate the average intensity on both sides of the dimming boundary.
The bottom 70 rows are selected and divided into three parts along
the slit direction. The areas selected are marked by dashed, dotted,
and solid boxes in the first and fourth panels in Figure \ref{fig4}.
The result is shown in Figure \ref{fig6}, using the same line styles as in
Figure \ref{fig4}. The intensity in the dimming region is 23\%
and 21\% less than that in the
pre-wave region for the \ion{Fe}{12} and \ion{Fe}{13} lines, respectively.
Furthermore, the intensity in the wave front region is 11\% higher than that in the
pre-wave region for the \ion{Fe}{13} line. This calculation for the
wave front is not applied to the \ion{Fe}{12} line, since
the wave front is not clear enough for this line.

\subsection{The Doppler Velocity}\label{vel}
In order to study the LOS velocity, the EIS slit tilt and the Hinode
orbital variation effects must be corrected. We use spline
fitting to correct the orbital variation. Note that those data points
that are during or close to the eclipse periods (i.e., columns 50--80,
160--190, and 270--300 in the data array) are ignored by setting their
weights to be zero manually. Since the active region is close to the
solar disk center, the projection effect is negligible.

The results show that there is a blue shift of more than 10 km s$^{-1}$
appearing in the dimming region. Note that, an uncertainty may exist in
the Doppler velocity due to the processing of the orbit variation, which
is much smaller than the blue shift. Additionally, similar to what we
have discussed above, before the eruption there is little mass motion in
the region where the dimmings are later observed (see Figure \ref{fig5},
right panel). Therefore, the existence of the outflows in the dimming
region is of physical significance. In some previous investigations on
EIT wave or eruption events, e.g., \citet{har03} and \citet{asa08}, line
splitting was observed in some coronal lines. These authors performed a
two-component Gaussian fitting and obtained a strong blue shift of over
100 km s$^{-1}$ for the moving component. For this event, we search the
line profiles at different positions and find no visible line splitting
except in a small region around the active region core. Thus, we only
make the single Gaussian fitting, which yields the outflow velocities
much smaller than those in \citet{har03} and \citet{asa08}.
Therefore, the velocity values obtained by these two methods cannot be
compared quantitatively.

Furthermore, we carefully investigate the spatial distribution of the
blue shift in Figure \ref{fig4}. We find that the outflows are generally confined in the
dimming region, and therefore are behind the EIT wave front, as shown in
the first and third panels of Figure \ref{fig4}. It is noted, however,
a patch of blue-shifted pattern, which is located around
$y \sim -82\pm4$\arcsec~and $x \sim -31$\arcsec, falls in the region with
density enhancement. We also find that,
different from the line intensity, the blue shift areas
for the \ion{Fe}{14} and \ion{Fe}{15} lines are similar to those for
the \ion{Fe}{12} and \ion{Fe}{13} lines.

\subsection{The Line Broadening}\label{broad}
Excess of line width over its thermal broadening is thought to be caused
by turbulent mass motions or waves. As shown in the middle column of
Figure \ref{fig3}, the line width significantly increases along the
boundary of the dimming region, seen as a ridge-like structure.
Comparing the result with that shown in Figure \ref{fig5}, it is
confirmed that the extra line broadening does not correspond to the
active region structures.

The width of the \ion{Fe}{12} $\lambda$195.12 line in the background
region is about 28 m\AA, while the line width in the ``ridge" is
generally larger than 30 m\AA. To obtain a more accurate result, 
we select several points in the ``ridge", the wave front, and
the pre-wave quiet region, as indicated by the quadrilaterals in the
second panel of Figure \ref{fig4}. Note that the position of the
middle box in this panel corresponds to the linear interpolation
between the wave fronts at 12:49:00 and 12:51:30 UT. The
result shows that the average line width in the ``ridge" is 31.3$\pm$0.9 m\AA,
which is 12\% broader than that in the quiet region, i.e., 28.0$\pm$0.6
m\AA. However, the widths for most of the points selected in the wave
front region, i.e., 28.9$\pm$0.5 m\AA, are almost the same as those in the
quiet region. The nonthermal velocities, $V_{non}$ for the \ion{Fe}{12} line,
the observed FWHM for the same line, FWHM$_{obs}$, and the instrumental FWHM,
FWHM$_{ins}$, are related by
\begin{equation}\label{non}
{\rm FWHM}_{obs}^{2}={\rm FWHM}_{ins}^{2}+4{\rm ln}2~\frac{\lambda^2}{c^2}~(\frac{2kT}{M}+V_{non}^2),
\end{equation}
where, the value of FWHM$_{ins}$ is 0.056 \AA, $\lambda$ is the wavelength (in \AA),
$c$ is the speed of light, $k$ is the Boltzmann constant, $T$ is the electron
temperature, and $M$ is the ion mass. The value of FWHM$_{obs}$ (in \AA)  can be obtained
from Gaussian fitting parameter.
The nonthermal velocities, converted from the line widths of the ``ridge", the wave
front, and the quiet region that are indicated in the second panel of Figure \ref{fig4}, are  $39.4\pm3.4$ km s$^{-1}$, $29.5\pm2.3$ km s$^{-1}$,
and $25.1\pm3.2$ km s$^{-1}$, respectively.

Furthermore, we investigate the spatial relationship between the line
broadening ``ridge" and the dimming region in Figure \ref{fig4}.
It is found that the location of the ``ridge" is generally
cospatial with the outer boundary of the dimming region. This may suggest that the
EIT wave-induced excess line broadening is the most significant at the
edge of the dimming region, while is generally negligible at
the EIT wave front. However, it is noticed that a patch of line broadening pattern, which is
located around $y \sim-82\pm4$\arcsec~and $x \sim -31$\arcsec, falls in the region with density enhancement. This
line broadening patch also presents blue shift as illustrated in
Section \ref{vel}. This seems to give us an impression that some part of
the EIT wave front may also present strong outflows and line broadenings. After
examining carefully the EUVI 195 {\AA} images in Figure 2 (the bottom row, mainly
the middle panel), we find that the density enhancement comes from
some bright loops embedded in the dimming region, rather than the EIT wave
front. This implies that the EIT wave front propagation, as denoted
by the red dotted line in the first panel of Figure 4, does not follow a strictly
straight line and may turn to be more vertical from ($x$, $y$)
$\sim$ ($-30$\arcsec, $-50$\arcsec). Note that a more
vertical line in the left panel of Figure 4 means a larger propagation
velocity. We also note that the width of the \ion{Fe}{13} $\lambda$202.04 line
shows a similar distribution to that of the \ion{Fe}{12} $\lambda$195.12
line, while the ``ridge" is not significant in the \ion{Fe}{14} and
\ion{Fe}{15} lines.

\section{Discussion}

EIT waves and the ensuing expanding dimmings are intriguing phenomena.
It was suggested that both of them result from the same process
\citep{chen05}. However, the nature of EIT waves is still under hot
debate. While many researchers consider EIT waves as fast-mode MHD waves
\citep{pom08,pat09,kie09,gopa09}, others believe that they are non-wave
perturbations \citep{del00, chen02, att07, del08, zhu09}. Spectroscopic
observations can help clarify the nature of EIT waves.

\citet{har01} discovered that there is a strong outflow in the dimming
through spectroscopic observations. Later, they found that there is
nearly no Doppler velocity at the EIT wave front \citep{har03}.
In this paper, we analyze the Hinode/EIS
observations of an EIT wave/dimming event. The results confirm the
Doppler velocity distribution found by \citet{har01,har03}.
These results are consistent with the model of
\citet{chen02}, which simultaneously interprets the formation of EIT
wave fronts and dimmings in terms of the stretching of magnetic field
lines, i.e., field lines are successively stretched to compress plasma
at the outer boundary to form bright fronts, with the bulging wake
forming the dimmings. We note that, the absence of Doppler velocity at the wave front
is also compatible with the fast-mode wave interpretation, since the velocity
disturbance induced by the wave front would be directed across the line of sight in
events observed near the disk center.

We also analyze the spectral line widths and find that they may
significantly increase at the outer edge of the dimming region.
The enhanced line widths associated
with outflows were reported both in the active region boundaries
\citep{dosc08} and CME-associated dimming regions \citep{mcin09}.
\citet{dosc08} explained the line broadening as caused by multiplicity of flows,
whereas \citet{mcin09} proposed an alternative model
that the broadening may be attributed to the growth of Alfv\'en wave
amplitudes in the rarified dimming region. It requires the convective
buffeting from the subsurface of the Sun to testify the Alfv\'en wave
hypothesis in the MHD modeling of CMEs. Instead, it is practical and
interesting to check whether the MHD numerical model of \citet{chen02,
chen05} can explain the spatial distribution of the EUV line broadening
in terms of multiplicity of flows as proposed by \citet{dosc08}.

\begin{figure}
\epsscale{1}
\plotone{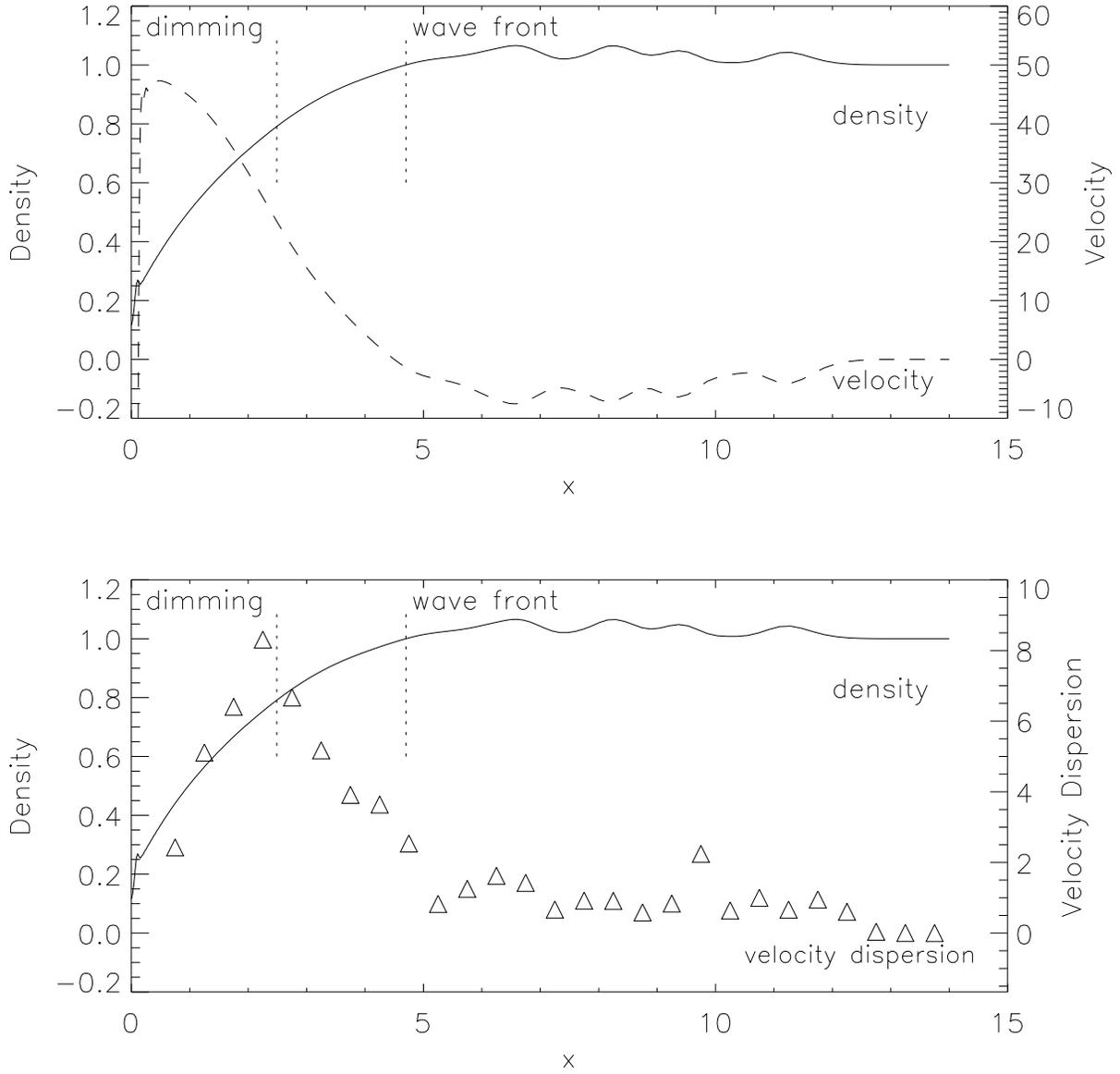}
\caption{Horizontal distribution of the density, LOS velocity,
velocity dispersion within each pixel at $y=4.5$ from a snapshot
of the
numerical model of \citet{chen02,chen05}. The vertical dotted
lines indicate the boundaries of the dimming region and the wave
front respectively, as defined in the text. All the physical units are arbitrary.
See text for details.}\label{fig7}
\end{figure}

We analyze the value at $y=4.5$ in the simulation result of
\citet{chen02,chen05}. The
temperature is approximately the quantity $T_{max}$ of the Fe XII $\lambda$195 line
as shown in Table 1 at
this height. We define those pixels with 0.8 times the pre-wave density or
less as the dimming region and pixels with density greater than the
pre-wave value as the wave front. The boundaries of the dimming and the
wave front are indicated respectively by the vertical dotted lines in
Figure \ref{fig7}. Note that, numerically, there is a smooth transition
between them.
The top panel of Figure \ref{fig7} shows the horizontal distributions of
plasma density and LOS velocity in the simulation results of
\citet{chen02, chen05}, which are consistent with our observations and
those of \citet{har01,har03}. In order to derive the spectral line width
distribution, we descretize the space in the simulation into pixels
according to the Hinode/EIS spatial resolution. We then use the
difference of the maximum and minimum values of the LOS velocities in
each ``pixel" to represent the physical effect that would lead
to the line broadening. As indicated by the bottom panel of Figure
\ref{fig7}, the result shows that the velocity dispersion within
each pixel reaches its maximum at the edge of the dimming region, but
becomes negligible at the wave front. This implies that the fieldline
stretching model of \citet{chen02, chen05}, which simultaneously
account for the formation of EIT waves and dimmings, can well explain
the observed distribution of line broadening in the dimming regions.

\acknowledgments
We thank the referee for constructive comments that helped
improve this manuscript. The research was supported by NSFC under grants 10828306 and
10933003 and by NKBRSF under grant 2006CB806302. Hinode is a
Japanese mission developed and launched by ISAS/JAXA,
collaborating with NAOJ as a domestic partner, NASA and STFC (UK)
as international partners. Scientific operation of the Hinode mission
is conducted by the Hinode science team organized at ISAS/JAXA.
This team mainly consists of scientists from institutes in the partner countries.
Support for the post-launch operation is provided by JAXA and NAOJ (Japan), STFC
(U.K.), NASA, ESA, and NSC (Norway). The SECCHI data used here were
produced by an international consortium of USA, UK, Germany, Belgium,
and France. We thank the STEREO/SECCHI consortium for providing
open access to their data.

\clearpage

\clearpage

\clearpage

\clearpage

\clearpage

\clearpage

\clearpage

\clearpage

\begin{deluxetable}{lrrr}
\tablecolumns{4}
\tablewidth{0pc}
\tablecaption{EIS lines used in this study}
\tablehead{
\colhead{Ion} & \colhead{~~~~~~~Wavelength (\AA)} & \colhead{~~~~Log T$_{max}$ (K)}
}
\startdata
\ion{Fe}{12}.... &195.12~~~~~~~~&6.1~~~~~~~~\\
\ion{Fe}{13}... &202.04~~~~~~~~&6.2~~~~~~~~\\
\ion{Fe}{14}...  &274.20~~~~~~~~&6.3~~~~~~~~\\
\ion{Fe}{15}....  &284.16~~~~~~~~&6.4~~~~~~~~\\
\enddata
\end{deluxetable}

\end{document}